\documentclass[letterpaper, 10 pt, conference]{ieeeconf}  % Comment this line out
\IEEEoverridecommandlockouts    
\usepackage{comment}
\usepackage{cite}
\newcommand{\eat}[1]{}
\usepackage{booktabs}
\usepackage{color}
\usepackage{xcolor, soul}
\usepackage [autostyle, english = american]{csquotes}
\MakeOuterQuote{"}
\usepackage{xcolor}
\let\labelindent\relax %Otherwise throw an error since IEEE also defines indent
\usepackage{enumitem}

\usepackage{multicol}
\usepackage{multirow}

\usepackage{mathtools}
    
\usepackage{amsmath}
\usepackage{amstext}% assumes amsmath package installed
\usepackage{amssymb}
\usepackage{amsfonts}
\usepackage{float}

\usepackage{amsthm}  %for makinf assumptions bold
\usepackage[normalem]{ulem} % for strikethrough
\usepackage{subfig}
\usepackage{caption}
\usepackage{todonotes}

\makeatletter

\newcommand{\Rmnum}[1]{\expandafter\@slowromancap\romannumeral #1@}
\usepackage{savesym}

\usepackage{algorithm}
\usepackage{algorithmicx} % <===========================================
\usepackage{algpseudocode} %
\savesymbol{AND}
\usepackage[group-separator={,},group-minimum-digits={3}]{siunitx}

\usepackage{graphicx} % for pdf, bitmapped graphics files
\usepackage{epsfig} % for postscript graphics files

\usepackage{times} % assumes new font selection scheme installed
\usepackage{amsmath} % assumes amsmath package installed
\usepackage{amssymb}  % assumes amsmath package installed
\usepackage{comment}
\makeatletter
\let\NAT@parse\undefined
\makeatother
\usepackage{hyperref}
\hypersetup{
   colorlinks=true,
    linkcolor= blue,
    allcolors=blue,
    citecolor = blue,
    filecolor=black,      
    urlcolor=blue,
    }
\usepackage{mathrsfs}

\usepackage[symbol]{footmisc}

\makeatletter
\renewcommand\@makefntext[1]{%
\setlength\parindent{1em}%
\noindent
\mbox{\@thefnmark}{#1}}
\makeatother

\begin{document}

\title{\LARGE \bf
MR-IDM - Merge Reactive Intelligent Driver Model: \\ Towards Enhancing Laterally Aware Car-following Models}
\author{ Dustin~Holley,
Jovin~D'sa,
Hossein~Nourkhiz Mahjoub,
Gibran~Ali,
Behdad~Chalaki,
Ehsan~Moradi-Pari
\thanks{\hangindent=0.5cm D. Holley is with the Global Center for Automotive Performance Simulation (email: dholley@gcaps.net)}
\thanks{J. D'sa, H. N. Mahjoub, B. Chalaki and E. Moradi-Pari are with Honda Research Institute, USA Inc. (email: jovin{\_}dsa@honda-ri.com)}
\thanks{G. Ali is with the Division of Data and Analytics, Virginia Tech Transportation Institute (email: gali@vtti.vt.edu)}
}

\maketitle
\thispagestyle{empty}
\pagestyle{empty}

\begin{abstract} 
This paper discusses the limitations of existing microscopic traffic models in accounting for the potential impacts of on-ramp vehicles on the car-following behavior of main-lane vehicles on highways.
We first surveyed U.S. on-ramps to choose a representative set of on-ramps and then collected real-world observational data from the merging vehicle's perspective in various traffic conditions ranging from free-flowing to rush-hour traffic jams.
Next, as our core contribution, we introduce a novel car-following model, called MR-IDM, for highway driving that reacts to merging vehicles in a realistic way. This proposed driving model can either be used in traffic simulators to generate realistic highway driving behavior or integrated into a prediction module for autonomous vehicles attempting to merge onto the highway. 
We quantitatively evaluated the effectiveness of our model and compared it against several other methods. We show that MR-IDM has the least error in mimicking the real-world data, while having features such as smoothness, stability, and lateral awareness.   

\end{abstract}
%%%%%%%%%%%%%%%%%%%%%%%%%%%%%%%%%%%%%%%%%%%%%%%%%%%%%%
\section{Introduction}
%\subsection{Motivation}
%\PARstart{S}{ince} road capacity has not increased at the same rate as population, traffic congestion in the U.S. has steadily grown from $1982$ to $2020$. In $2019$, traffic congestion in the United States caused drivers to spend an additional $8.7$ billion hours on the road and purchase an additional $3.5$ billion gallons of gasoline, totaling \$$190$ billion in congestion costs \cite{Schrank2021}. 
\PARstart{D}{ue} to limited road capacity compared to population growth, traffic congestion in the U.S. has steadily risen from 1982 to 2020. In 2019 alone, congestion resulted in an extra 8.7 billion hours on the road and 3.5 billion additional gallons of gasoline, costing a total of $190$ billion \cite{Schrank2021}. Apart from the adverse  environmental impact, increased congestion has also been shown to increase driver stress, anxiety, and aggression \cite{hennessy1999traffic,zinzow2018driving}. The highway merging scenario, which consists of on-ramps, off-ramps, and interchanges, is of particular importance because it accounts for some of the most significant causes of congestion on U.S. roadways\cite{margiotta2011agency}. 

%\todoall{talk about negotation, high speed differential, limited time/space to complete maneuver etc.}  
%Autonomous vehicles (AVs) have the potential to reduce traffic congestion and improve safety in several different driving scenarios, including highway merging. 
Autonomous vehicles (AVs) can alleviate traffic congestion and enhance safety in various driving scenarios, including highway merging. A survey of different studies that consider the impact of AVs on improving traffic congestion in highway merging scenarios can be found in \cite{rios2016survey} and \cite{li2013survey}. 
Apart from being a source of congestion, highway merging is also a challenging driving scenario, as the driver in the merge lane needs to negotiate with other drivers in the main lane and make several key decisions with hard constraints on time and road geometry \cite{zgonnikov2020should}. 
 Both academia and the automotive industry have shown a strong interest in the development of AV technology to reduce driver anxiety during highway merges. However, testing and validating AV technology for applications such as highway merging requires a simulation environment with a driver model that can represent the dynamics of human drivers in main-lane traffic, as the interaction between AVs and human-driven vehicles is crucial in such driving scenarios.
 Specifically, the dynamics of the yielding behavior for main-lane traffic participants is of special importance since it has a direct impact on the decision-making and control algorithms utilized by AVs attempting to merge onto the highway.

%\todoall{justification of why we did not use exisitng simualtion packages and why our work is important}
%Eventhough several traffic simulation software exist such as VISSIM, SUMO as well as some more high fidelity dynamics simulator such as IPG CARMAKER, there is no readily available simulation tool to simulate freeway merging. Even if one can create the freeway geometries in these simulation software the traffic behavior in these commercial software tend be mainly 1D in nature due to the use of car-following models such as IDM, GM model etc. Some software such as VISSIM have a lane change decision making model embedded such as MOBIL which work well to simulate lane change decisions during general freeway driving but are not specically tailored for highway merges which are a form of forced merge scenes. Thus in this work we have tried to propose car-following models that have the ability to react to a forced merge situations and the model has been validated to work well in merging situations. 

\subsection{Related Work}
Over the past few decades, a number of articles have examined microscopic models to characterize the driving behavior of traffic participants in various scenarios. These methods can be divided into two main categories: car-following and lane-changing models. Car-following models, such as the Gipps model \cite{gipps1981behavioural}, Newell model \cite{newell2002simplified}, GM model\cite{gazis1961nonlinear}, optimal velocity model\cite{bando1995dynamical}, and intelligent driver model (IDM) \cite{treiber2000congested}, primarily characterize longitudinal behavior. Lane-changing models, as their name suggests, focus on developing models to explain the lateral behavior of drivers. A thorough review of the main car-following models is provided in \cite{matcha2020simulation,zhang2023review}. 

%% IDM 
We focus our attention on IDM due to its wide applications in the AV community and intuitive parameters. IDM is a car-following model describing the longitudinal motion of a vehicle while reacting to the stimulus from the preceding vehicles on the same lane \cite{treiber2000congested}. Since its original formulation, there have been several attempts to improve the model. Treiber et al. extended the original IDM to consider several vehicles ahead \cite{treiber2006delays}, while Zong et al. \cite{zong2021improved} considered the effects of both multiple rear and front vehicles. Additionally, Kesting et al.\cite{kesting2010enhanced} presented an enhanced IDM model to eliminate some of the unrealistic behavior, such as overreacting to the preceding vehicles in cut-in situations, which greatly improved the model's smoothness and stability. %However, as authors in \cite{albeaik2022limitations} suggested, for certain class of initial conditions, IDM still shows some undesirable car-following behavior.  

Another area of research in developing car-following models considers the role of visual cues for the assessment of the urgency of the driver on deceleration behavior. The idea of visual looming is that an object moving towards the driver causes the driver to perceive an upcoming collision and react to it \cite{lee1976theory,liebermann1995field}. This looming angle, i.e., the angle induced by a moving object on the retina of the driver of the host car, as well as the rate of change of looming angle, have been shown to have an impact on the braking behavior of the driver \cite{terry2008role,xue2018using}.

To date, numerous research articles published in the literature have concentrated on developing models to characterize the driving behavior of vehicles attempting to merge onto a highway \cite{kondyli2011modeling}, but few studies address modeling main-lane vehicles reacting to vehicles in the merging lane \cite{wan2014modeling,ward2017probabilistic}. Wan et al. \cite{wan2014modeling} extended a simple car-following model for main-lane vehicles at highway weaving sections to also react to the lateral vehicles aimed at merging through a looming angle. Ward et al. \cite{ward2017probabilistic} employed IDM as a prediction model in their motion planner for highway merging to determine main-lane vehicles' reaction to the merging vehicle by defining a critical path section beyond which the ego car in the merge lane becomes the leader for the car on the main lane.

%\todoall{Listing the papers which have considered mainlane actor model that react to the merge lane
%1. 1-D model (idm), shortcoming of it 
%2. any 2-d models, what was the pros and cons
%3. Try to connect it to our proposed approach in this paper}

%\todoall{Some related paper on autonomous driving on merge scenario and also traffic modeling and simulations, mentioning the shortages in the literature
%paragraph \cite{treiber2000congested}IDM ,
%Many car-following models exist such as GM, Gipps, IDM. IDM is very well known and widely studied to show good propertiestasl about IDM enhancement:0 acceleration heuristic since used in the modelling \cite{Kesting_2010}.
%}
\subsection{Contributions of the Paper}

% In order to perform high fidelity testing of CAV technology, it is important to not only test the new technology in isolation but also to test its performance in a integrated fashion along with other Despite the existence of numerous commercial simulation softwares such as such as CarSim and IPG-CarMaker which provide high-fidelity vehicle dynamic , microscopic traffic simulators such as PTV-VISSIM and SUMO as well as some  the traffic behavior for remote cars in the scene in these commercial software tends to be different extensions of car-following models like IDM primarily one-dimensional in character. 
High-fidelity vehicle dynamics software, such as CarSim \cite{manual2002mechanical}, IPG-CarMaker \cite{carmaker2021reference}, etc., are widely used in the industry to perform large-scale testing of different AV technologies to check for safety-critical compliance of integrated systems in different real-world conditions. However, most of these software use different car-following models, such as some extensions of IDM or the Wiedemann $99$ \cite{Wiedemann99} model, which do not generalize particularly well to highway merge dynamics. Although some microscopic traffic simulators such as VISSIM \cite{ptv2018ptv}, have an integrated lane-changing model to simulate both mandatory and discretionary lane changes to mimic the lane-change decisions of vehicles during regular highway driving %\cite{ptv2018ptv} \cite{lopez2018microscopic} 
, they do not characterize the driving behavior of highway drivers reacting to the on-ramp traffic in forced-merge circumstances, which can be seen in real-world scenarios. 

%\todoall{ 1. Introducing the learnings from a new dataset specifically tailored for on-ramp merging?2. Enhancing existing driver behavior models3. eventhough LC decision making models exist, they are mainly focused on general LC on freeways rather than specifically targetting Merge scenarios. G: unique data set it also includes denser, also we did site selection which is a good representation of highway in the us, we have many events and many sites.Site selection is very important and unique aspect of this work. }

%\todoall{todo: Summarizing the work here in one paragraph}

To the best of our knowledge, this is one of the first attempts to develop traffic models based on extensive collected data from different highway on-ramps in the U.S. to characterize the driving behavior of highway drivers that are also reacting to the on-ramp traffic. 
We believe that this paper advances the state of the art in the following ways. First, we performed a highway on-ramp survey in the U.S. to identify a set of representative candidate on-ramps, including $16$ different sites that would provide a wide range of examples. Second, not only did we collect on-site data from multiple sources to avoid potential limitations in aerial datasets such as NGSIM 
 \cite{kovvali2007video} as mentioned in \cite{coifman2017critical}, we also covered the full spectrum of traffic conditions, rather than just focusing on moderately congested traffic. Finally, our novel car-following model is also capable of reacting to merging vehicles in a realistic manner. The significance of our work is twofold: not only can our driving model be utilized in traffic simulators to generate more realistic highway driving behavior, but it can also be integrated in a prediction module for AVs attempting to merge onto the highway.

%\todoall{Defining abbreviations somewhere here}

%todoall{Maybe a couple of points to compare our }
\subsection{Organization of the Paper}
The structure of the paper is as follows. In Section \ref{sec:site}, we introduce details of the real-world dataset that was collected as part of this study, including information about the methodology of site selection, different data processing steps, and categorization of the data. In Section \ref{sec:simulDev}, we introduce our modeling framework by first covering some existing models that we investigate in this work, and then we provide the formulations for our recently developed models. 
In Section \ref{sec:behaveModel}, we demonstrate the performance of our models to replicate real-world behavior, and we provide concluding remarks and future discussions in Section  \ref{sec:conclusion}.

%\subsection{Preliminaries}
%\todoall{All the materials below are just copied from the report, they need to be revised and polished. }
\section{Data Collection and Processing}\label{sec:site}
%Naturalistic driving data was collected with a host vehicle on 16 different on-ramps (sites) that were chosen for their potential to produce congested traffic conditions. The data was collected from lidar, camera, GPS, inertial measurement unit, and Mobileye sensors, as well as the host vehicle’s controller area network bus channels. This data was then segmented into merge events and categorized into one of four scenario types . The LiDAR data was processed to find actor bounding boxes and their centroids, and further processing provided lane number assignments and in-lane positions. The resultant traffic data was then processed to combine actor IDs representing a single actor, interpolated to generate time-synchronous time histories, and smoothed to remove noise and fill gaps.

\begin{figure}
    \centering
\includegraphics[width=0.95\linewidth]{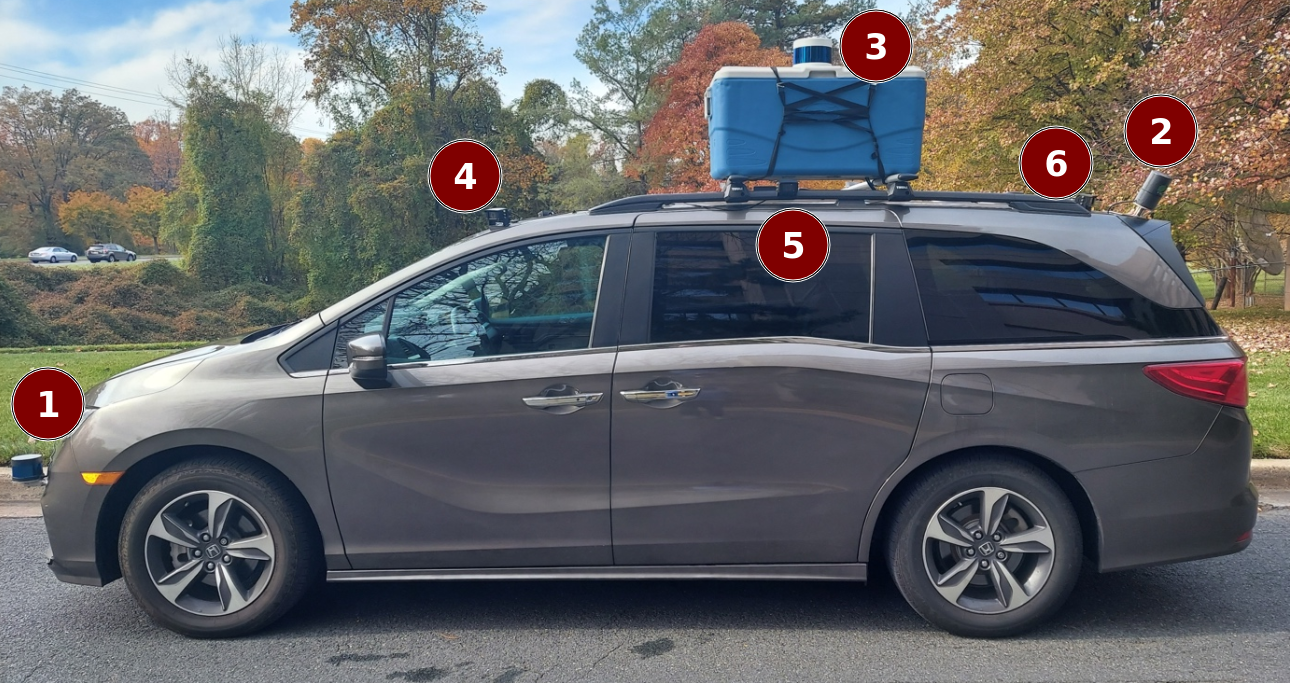}  
\caption{The data collection vehicle showing the location of three lidars and the three cameras visible from the side view.}
    \label{fig_vehicle_sensors}
\end{figure}

\begin{figure}
    \centering
\includegraphics[width=0.95\linewidth]{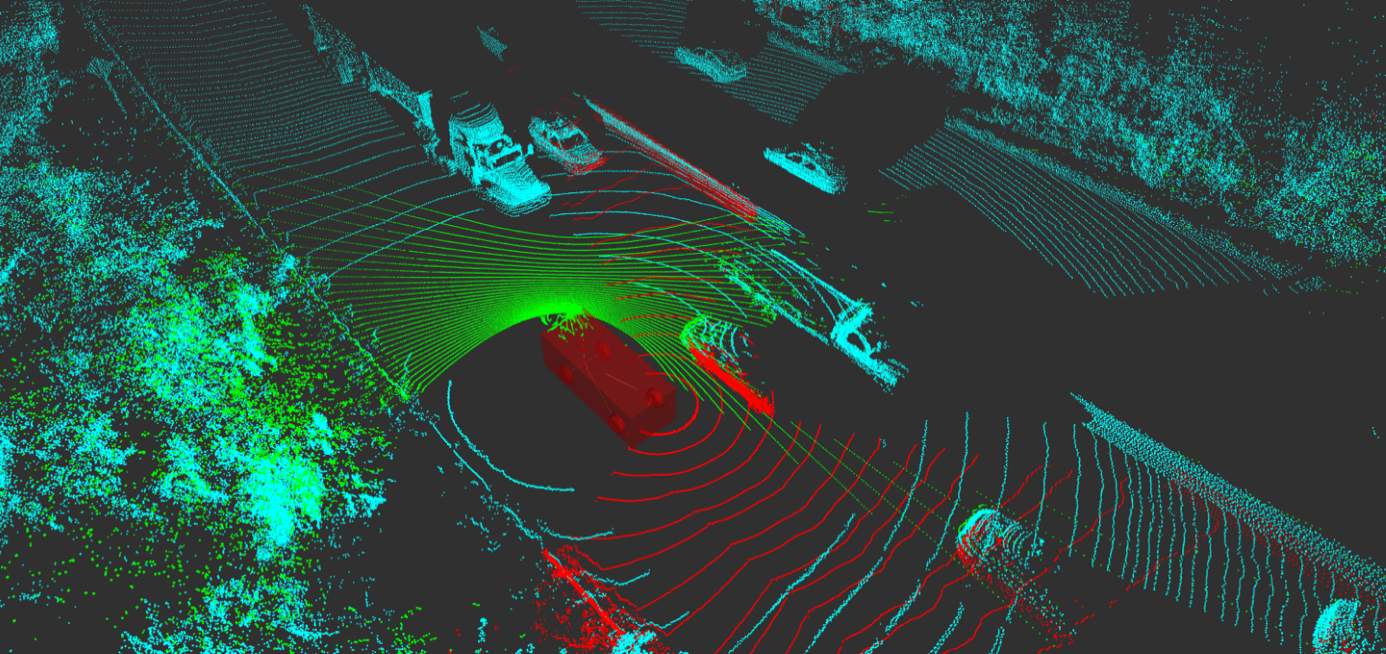}  
\caption{Point clouds collected from the three lidars; red shows front lidar point cloud, green shows tilted rear lidar point cloud, and cyan shows camouflaged top lidar point cloud.}
    \label{fig_lidar_spread}
\end{figure}

The data collection used in this study was designed to collect real-world data from a merging vehicle perspective at a diverse set of on-ramps in various traffic conditions.  To capture the data, a $2018$ Honda Odyssey was instrumented with three lidars, eight cameras, a Mobileye, and a high-precision navigation system. In addition to these sensors, data was also recorded from the various onboard sensors using the Controller Area Network (CAN) bus. Collectively, these sensors provided high-precision data about the vehicle location, kinematics, and surrounding traffic. Fig. \ref{fig_vehicle_sensors} shows the location of the various sensors with front lidar (1), tilted rear lidar for improved visibility in close merging situations (2), top lidar camouflaged with a cooler (3), and three of the eight cameras visible in the side view (4-6).  Fig. \ref{fig_lidar_spread} illustrates that the point clouds of the three lidars capture all the relevant traffic interactions. 

\begin{figure}
    \centering
   % \begin{subfigure}
        \includegraphics[width=\linewidth]{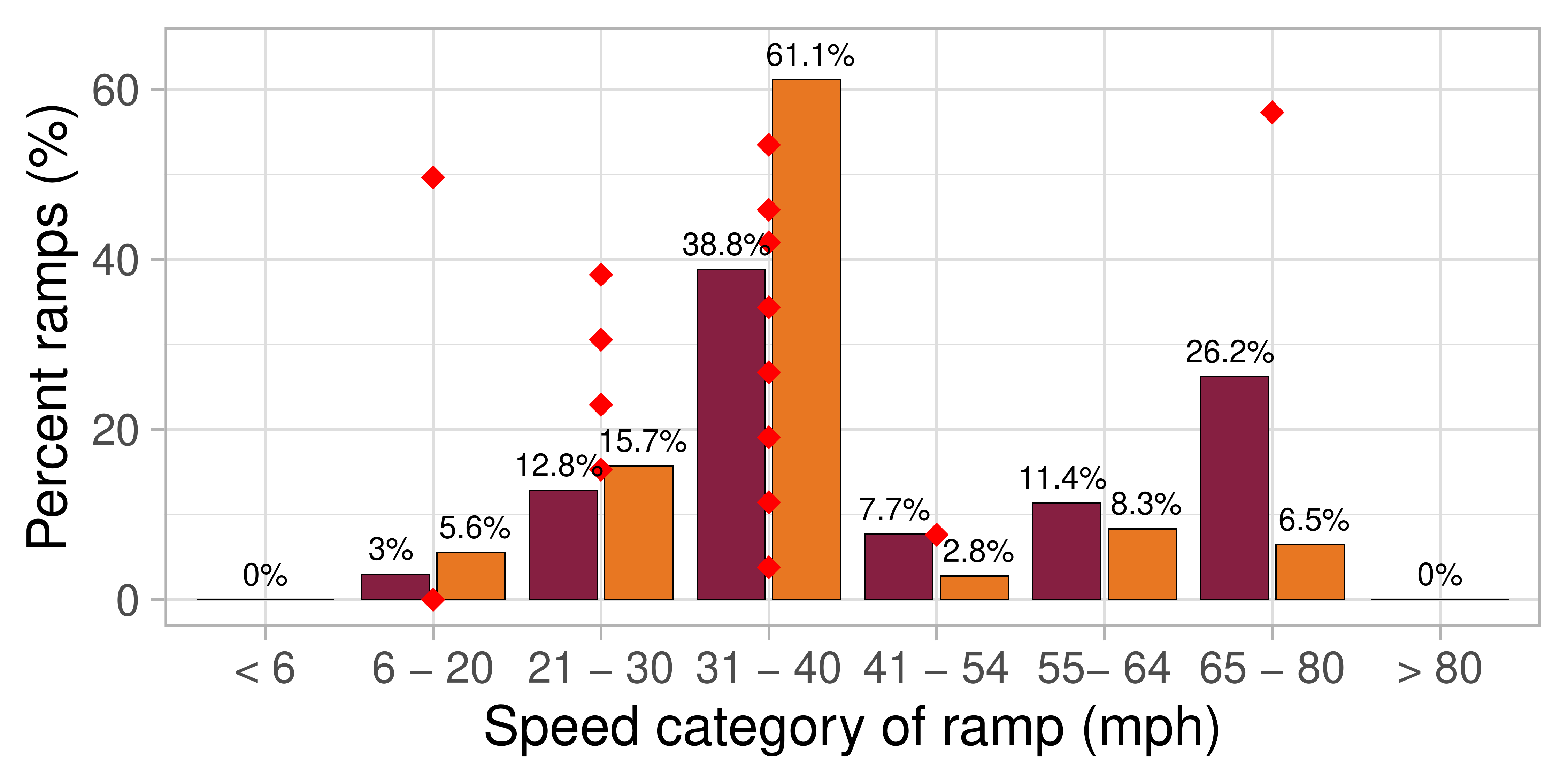}  
        \label{fig_site_selection_rmp_speed_cat}
    %\end{subfigure}
    %\begin{subfigure}
        \includegraphics[width=\linewidth]{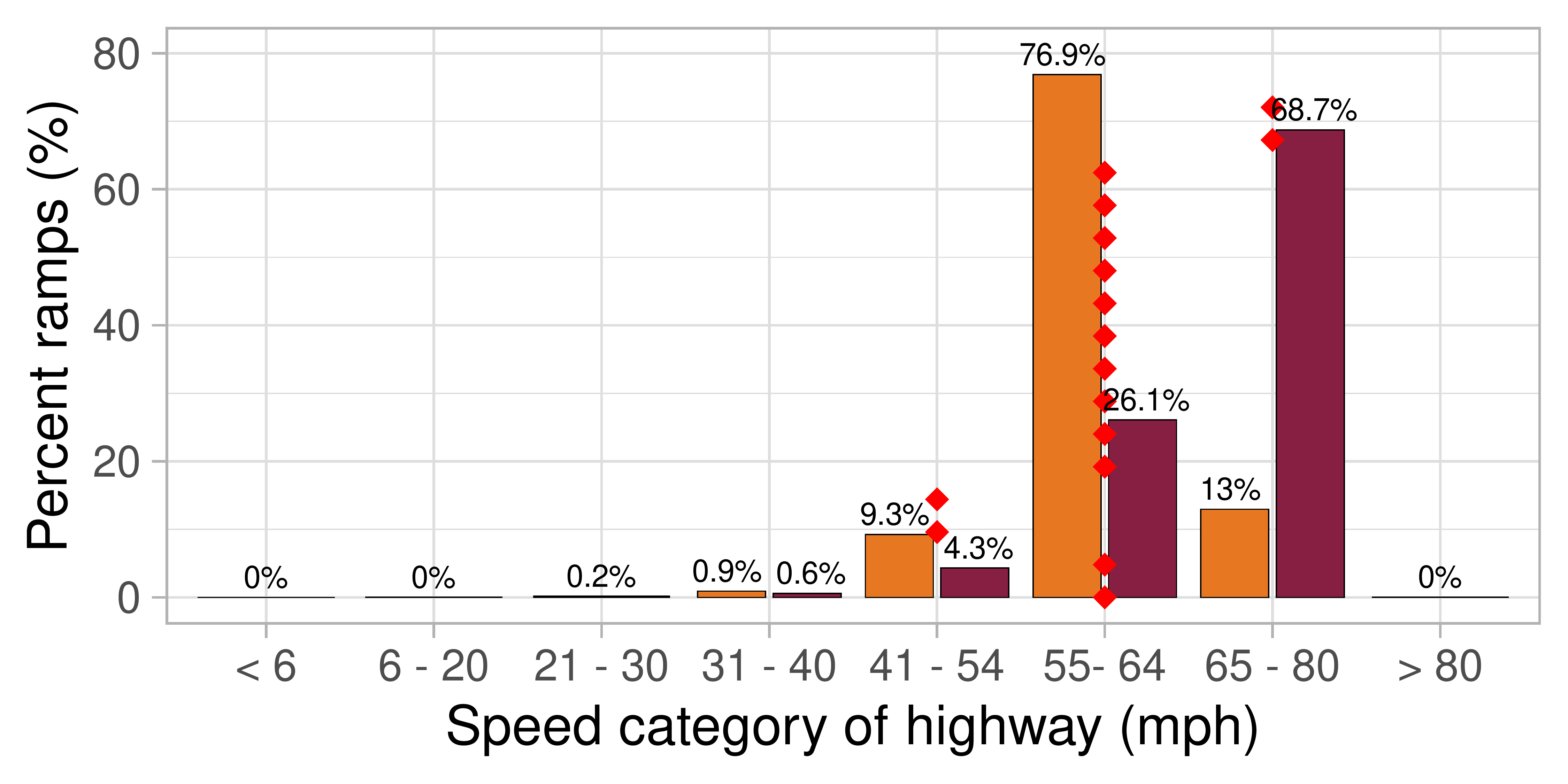}  
        \label{fig_site_selection_hw_speed_cat}
    %\end{subfigure}
    %\begin{subfigure}
        \includegraphics[width=\linewidth]{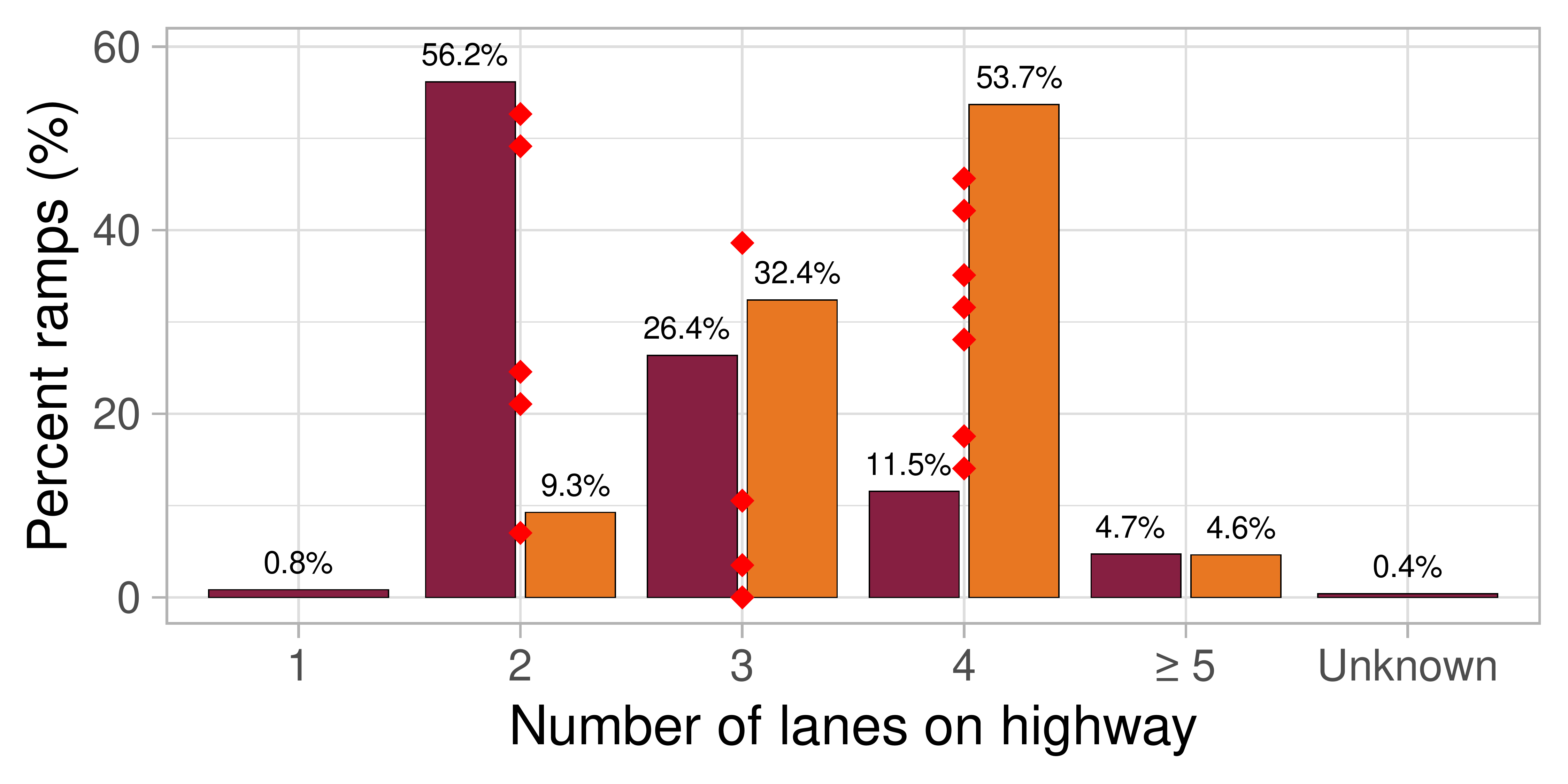}  
        \label{fig_site_selection_num_lanes}
    %\end{subfigure}
    %\begin{subfigure}
        \includegraphics[width=\linewidth]{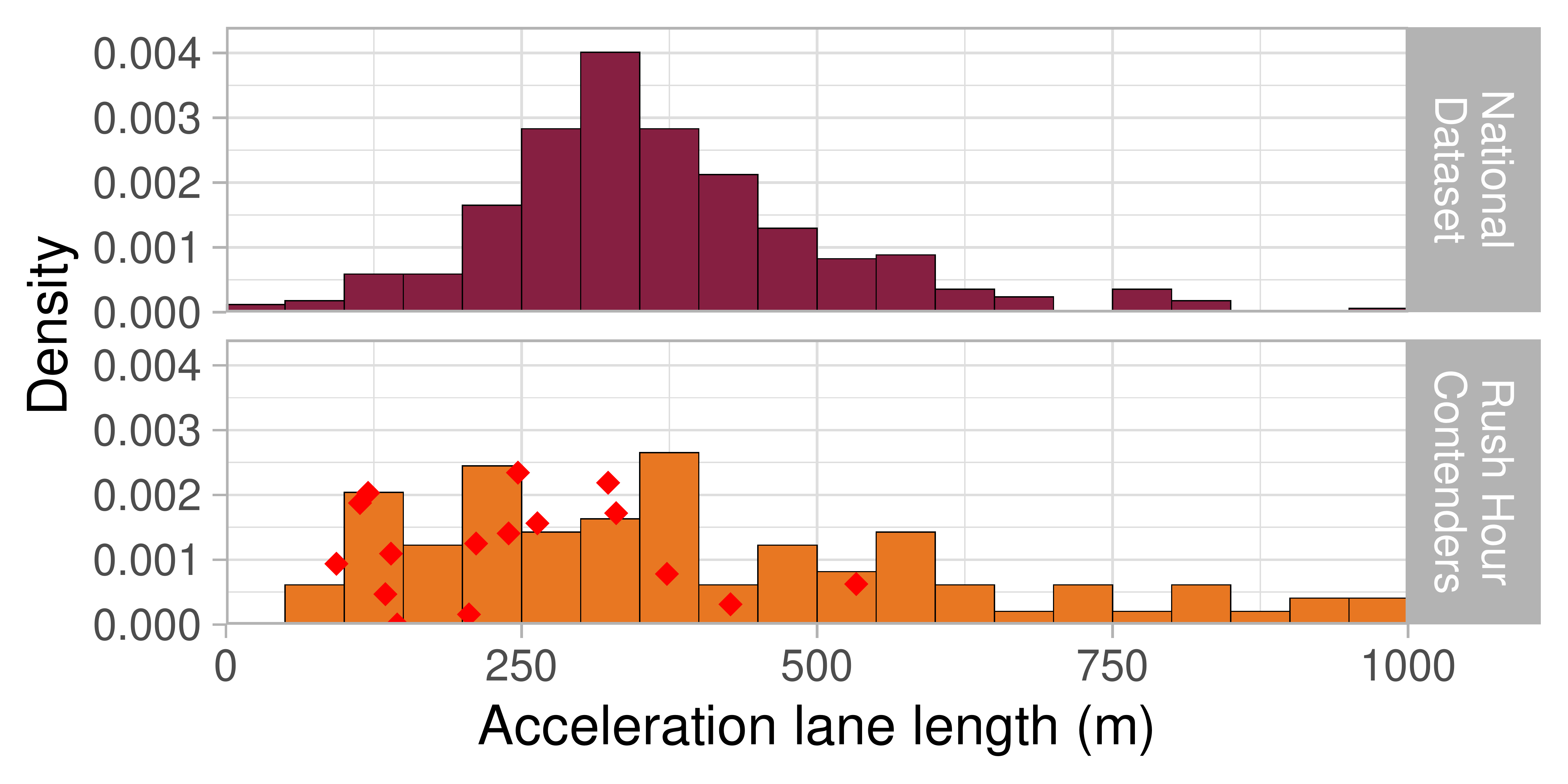}  
        \label{fig_site_selection_lane_lngth}
    %\end{subfigure}
    %\begin{subfigure}
        \includegraphics[width=\linewidth]{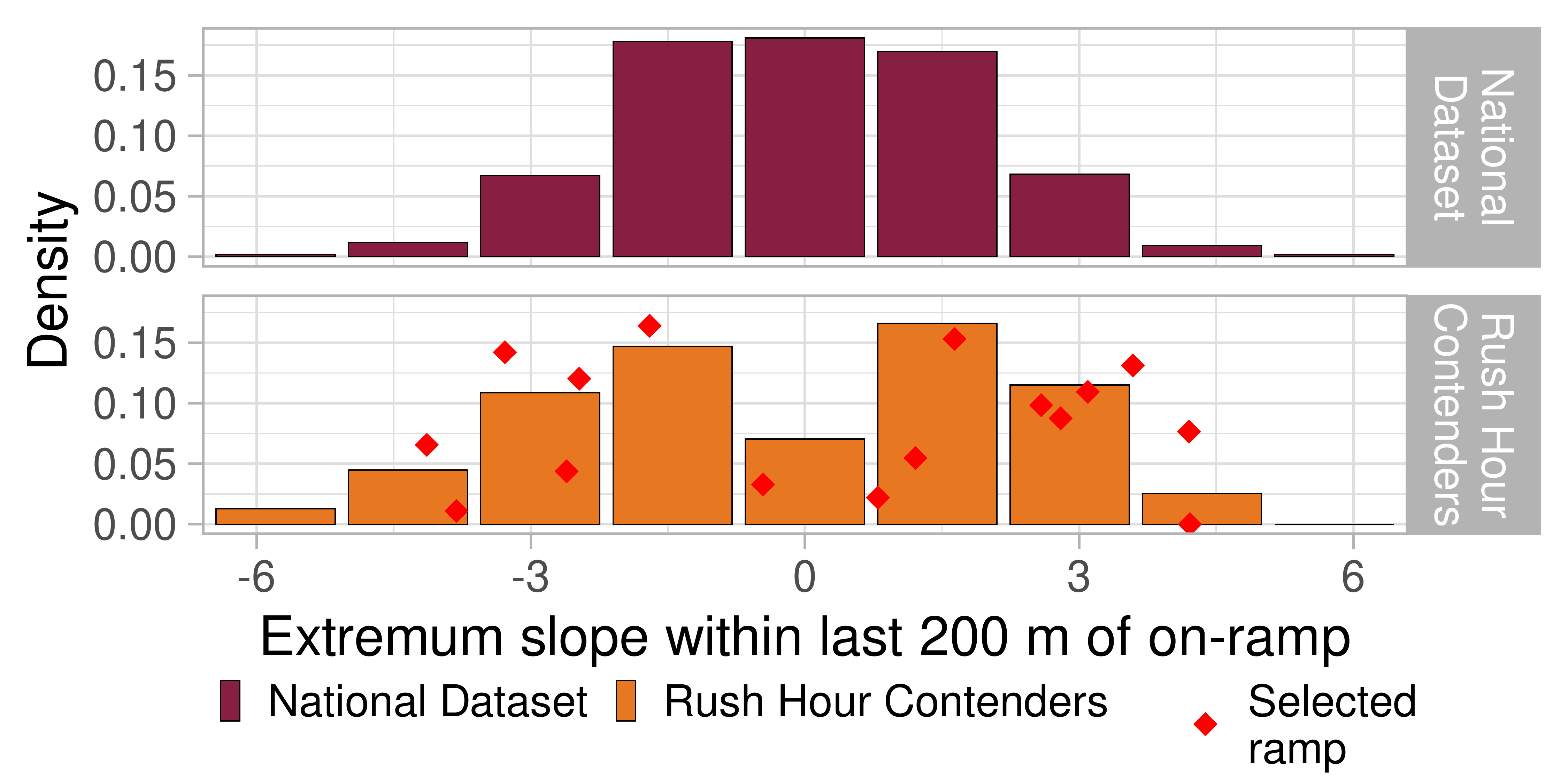}  
        \label{fig_site_selection_slope}
    %\end{subfigure}
    \caption{The distribution of national on-ramps, rush hour contenders, and selected data collection sites for speed category of on-ramp, speed category of highway, number of lanes on highway, acceleration lane length, and the on-ramp slope.}
    \label{fig_site_selection_stats}
\end{figure}

To capture meaningful insights about traffic around merges, it was important to collect data from a diverse set of on-ramps under varied traffic conditions. For this purpose, the traffic trends of on-ramps within a 5-hour drive from Blacksburg, Virginia, were analyzed and about one hundred potential candidates were identified that experienced slow rush-hour traffic. Data collection sites were chosen from this list after a statistical analysis of various on-ramp parameters. Fig. \ref{fig_site_selection_stats} shows the distribution of on-ramp speed category, highway speed category, number of highway lanes, acceleration lane length, and extremum slope of the on-ramp for the rush hour candidates, as well as the set of all on-ramps in the U.S.  \cite{here_technologies_guide_nodate}. Seven data collection sites, consisting of 16 on-ramps, were chosen. These on-ramps are shown as red-colored diamond marks on Fig. \ref{fig_site_selection_stats}, with the position along the x-axis representing the value of the metric, and the position along the y-axis being randomly assigned to avoid overlap. These selected sites cover $80$\%-$90$\% of all U.S. on-ramps in terms of the above-listed metrics.  The design of the data collection makes this dataset representative of most U.S. on-ramps in terms of important ramp parameters as well as various traffic conditions. This is a major improvement over older datasets used for modeling vehicle interactions, which included only one or two merging areas. The dataset records notable variations in driving behavior, including merging patterns, compliance with traffic regulations, and driver aggression, which frequently distinguish driving habits across different regions. Nevertheless, in order to comprehensively capture all regional driving behaviors, data collection would need to encompass the entire United States, which was beyond the project's scope.

\begin{figure}
    \centering
\includegraphics[width=\linewidth]{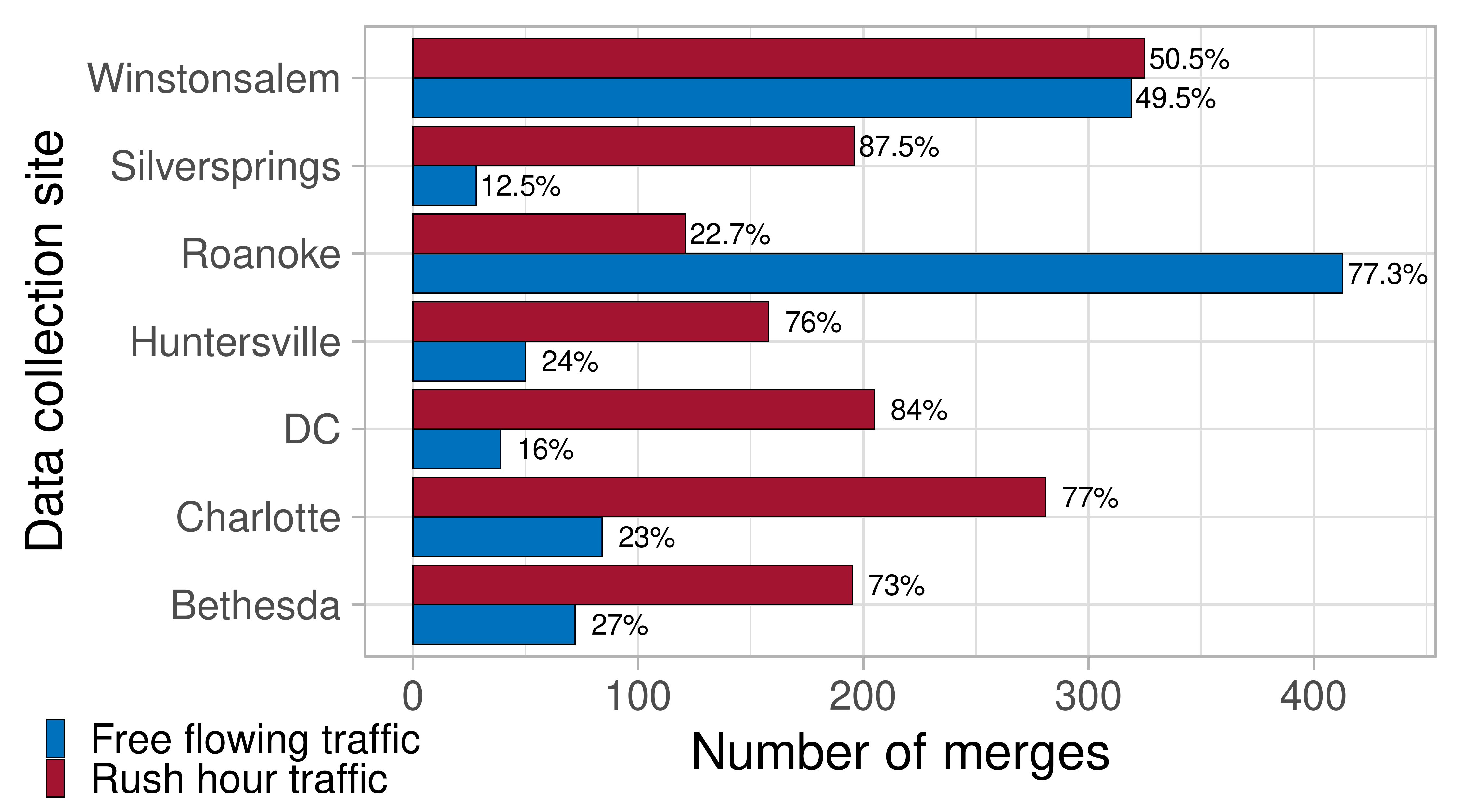}  
\caption{Number of merges collected in various traffic conditions at the different data collection sites.}
    \label{cases_by_sites}
\end{figure}

An expert driver drove the ego vehicle at each site for multiple days and accumulated over $2,400$ total merges between July and November $2021$. The driver received instructions to drive in a normal manner, placing a priority on safety and refraining from influencing other road users to engage in unusual behaviors.  Fig. \ref{cases_by_sites} shows the total number of merges collected at each site and the relative  proportion of free-flowing versus rush-hour traffic cases.
Once the data collection was completed at a site, the data was transferred to computing clusters where various data processing steps were completed. These steps included processing ego vehicle kinematics, traffic actor properties, and extracting other metrics used in data analysis and traffic modelling.

%\subsection{Actor Definitions}
As shown in Fig. \ref{actor definitions}, we define the various traffic participants as follows:

\begin{figure}
    \centering
\includegraphics[width=0.95\linewidth]{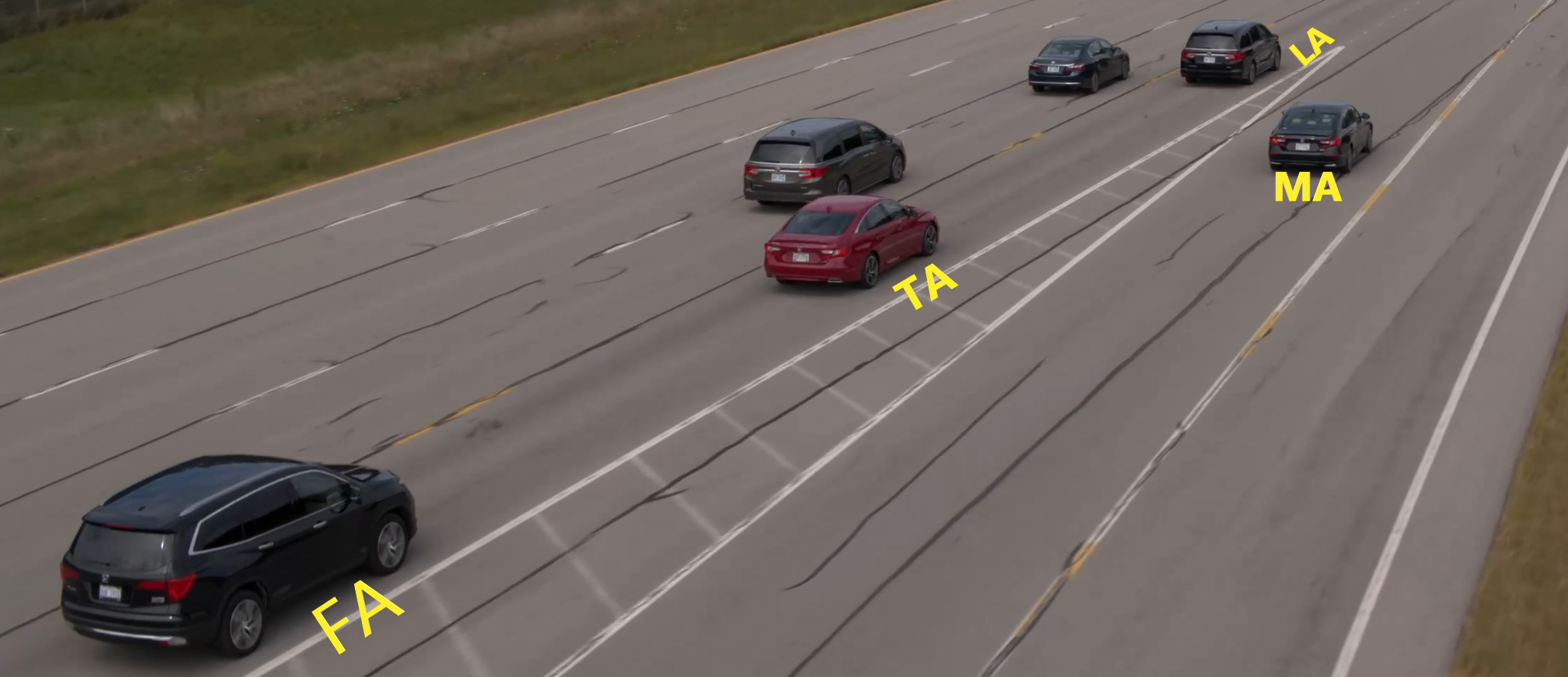}  
\caption{Actor definitions in traffic behavior modeling setup. 
TA: main traffic actor that is interacting with merging car; MA: merging actor; FA: following actor; LA: lead actor.}
    \label{actor definitions}
\end{figure}

\begin{itemize}
    \item Merge Actor (MA): vehicle merging onto the highway.
    \item Traffic Actor (TA): vehicle of interest in the main highway lane that is performing the yielding maneuver directly in response to the merge actor, MA.
    \item Lead Actor (LA): vehicle directly ahead of TA.
    \item Following Actor (FA): vehicle directly behind TA.
\end{itemize}

The collected events were filtered for validity based on the following:
\begin{itemize}
    \item Events must have at least 10 seconds of data.
    \item Events must have a TA, MA, and LA.
    \item The only lane change must be the MA’s merge, since the targeted models are only designed for car-following and merge-interaction behavior.
\end{itemize}

After the successful filtering and optimization of events, 401 merge events were used for model analysis.

%\todoall{I've kept the following subsection titles from the "PMToo" draft for now to decide later if we want to keep them in this paper or no}
%\subsection{Site Selection}

%\subsection{Site-based Data Analysis}

\section{Modeling Framework}\label{sec:simulDev}

 One goal of this work was to develop a traffic model that can be integrated into a high-fidelity simulation environment like IPG-CarMaker, which would allow testing of various AV motion planning and control algorithms. To achieve this, we identify the following key model requirements:

 %of a test car of is to be able to realistically simulate the reaction of highway traffic to a merging car in a advantages of finding a traffic behavior model which is able to realistically react to the merging traffic on the ramps is its utilization in high-fidelity simulator platforms to make them legit environments for evaluation of merge-related automated system designs.However, in order to achieve this the model needs to meet some key criteria as listed here:

 {\begin{itemize}
    \item Accuracy:  Model needs to have high accuracy in terms of its similarities to the reactive behavior dynamics in real-world data.
    \item Computational Efficiency: Model should be easily scalable.
    \item Tunability: Model should be easily customized to simulate different behaviors. 
    \item Interpretability: Model needs to have parameters with interpretable (preferably physical) meanings to allow easy set up of different test scenarios with specific requirements (preferably no black-box solution).
\end{itemize}}
 
 Having these requirements in mind, in this section we present our proposed models for highway traffic modeling at merging scenarios. %We will discuss their performance evaluation and comparison results later in the subsequent sections of the paper.

\subsection{ Proposed Models}

We first investigated the well-known IDM \cite{treiber2000congested} and looming angle \cite{wan2014modeling} models to determine if they can provide good correlation to the reactive behavior to highway merging events. Since these models did not show good correlation to the real-world data, we then investigated new model design variations as explained in the following sections.

\subsubsection{IDM-MSA}
One potential influence on the TA's reactivity to the MA is the longitudinal proximity of the MA to the end of the merge ramp. The closer the MA is to the end of the ramp, the more urgent the merge interaction becomes. The IDM-MSA (MOBIL-Scaled Acceleration) model uses this increasing proximity to transition the TA's leader from the LA to the MA.

We employ the IDM car-following model as follows:

\begin{equation}
s^* = s_0+vT+\frac{v(v-v_l)}{2\sqrt{ab}},
\end{equation}

\begin{equation}
\dot{v} = a(1-(\frac{v}{v_0})^4-(\frac{s^*}{s})^2),
\end{equation}
where $s^{*}$ is the desired distance gap between the TA and the LA, $s$ is their current distance gap, $v$ is the TA's speed, $v_l$ is the LA's speed, $s_0$ is the minimum distance gap, $T$ is the desired headway, $v_0$ is the desired speed, $a$ is the maximum acceleration, and $b$ is the comfort deceleration.

In IDM-MSA, we alter IDM's acceleration equation gap term as follows:
\begin{equation}
(\frac{s^*}{s})^2 \rightarrow r(\frac{s^*_{\textit{MA}}}{s_{\textit{MA}}})^2 + (1-r)(\frac{s^*_{\textit{LA}}}{s_{\textit{LA}}})^2,
\end{equation}
where $r$ is the scaling factor, $s^*_{\textit{MA}}$ is the IDM desired gap to the MA, $s_{\textit{MA}}$ is the current gap to the MA, $s^*_{\textit{LA}}$ is the IDM desired gap to the LA, and $s_{\textit{LA}}$ is the current gap to the LA.

We utilized the incentive criterion inequality from the MOBIL lane-change model \cite{kesting2007general} in order to compute $r$, deriving the scaling factor by computing the percentage of the lane-change threshold currently reached by the MA:

%To calculate $r$, the incentive criterion inequality from the MOBIL lane-change model \cite{kesting2007general} was used. The scaling factor was derived by computing the percentage of the lane-change threshold currently reached by MA:
\begin{equation}
r = \frac{\Delta a_{\textit{MA}} + p(\Delta a_{\textit{TA}})}{\Delta a_{th}},
\end{equation}
where $\Delta a_{\textit{MA}}$ is the change in the MA's acceleration induced by the lane change (with the MA's leader changing from the ramp end to the LA), $\Delta a_{\textit{TA}}$ is the change in the TA's acceleration (with the TA's leader changing from the LA to the MA), $p$ is the MOBIL politeness factor parameter, $\Delta a_{th}$ is the decision threshold parameter, and $r$ is the bound between $0$ and $1$.

\subsubsection{Looming-Mod}\label{sec:looming-mod}
The looming study's \cite{wan2014modeling} putative follower acceleration model contains two components that can produce unwanted acceleration during merge interactions. Looming-Mod attempts to address these issues by modifying the looming model.

The first component is the transition between two states: from (1) the TA reacting to both the MA and LA to (2) the TA reacting to just the MA. At this transition, the LA's influence on the TA is suddenly removed, which can result in a jump in positive acceleration. Since the simulated merge period occurs up until the MA crosses into the TA's lane, the LA can be assumed to be relevant throughout the merge interaction. Thus, the Looming-Mod does not include the second state.

The second component is the effect of the looming model's $\dot{\theta}_{M}$ on acceleration. This variable is positive when the MA is faster than the TA and negative when the MA is slower. Thus, this variable will only contribute to negative acceleration if the MA is slower than the TA, and can contribute positive acceleration if the MA is attempting to pass the TA, effectively closing off any potential merge gap. To prevent aggressive positive acceleration in response to a passing TA, Looming-Mod replaces $\dot{\theta}_{M}$ with $\min([0,\dot{\theta}_{M}])$.

\subsubsection{IDM-Looming}
One major difference between IDM-based models and the looming-based models is their intended operational domain. The IDM was developed for car-following situations, while the looming model was developed for merge negotiations. Since both models show stronger performance within their respective domains, we developed a new model, IDM-Looming. Here we utilize the model proposed by by Kesting et al. \cite{kesting2010enhanced} which we refer to as IDM with Constant Acceleration Heuristic (IDM-CAH) when the TA is purely car-following and incorporates Looming-Mod during merge interactions.

IDM-CAH reduces unrealistic IDM deceleration in response to smaller-than-desired gaps, such as the moment when the MA takes over for the LA as the leader. Given distance gap $s$, TA speed $v$, leader speed $v_l$, and leader acceleration $a_l$, we derive the new acceleration $\dot{v}$ from the following equations:
\begin{equation}
\dot{v}_{\textit{CAH}} = 
\left\{ 
  \begin{array}{ l l }
    \dfrac{v^2\tilde{a}_{l}}{v_l^2-2s\tilde{a}_{l}}, & \: \textrm{if } v_l(v-v_l) \le -2s\tilde{a}_{l} \\ \\
    \tilde{a}_{l}-\frac{(v-v_l)^2H(v-v_l)}{2s},     & \: \textrm{otherwise}
  \end{array}
\right.
\end{equation}
where $\tilde{a}_{l} = \min(a_l,a)$, and

\begin{equation}
\dot{v} = 
\left\{ 
  \begin{array}{ c l }
    \dot{v}_{\textit{IDM}},& \: \textrm{if } \dot{v}_{\textit{IDM}} \ge \dot{v}_{\textit{CAH}} \\
    \\
    (1-c)\dot{v}_{\textit{IDM}} \\ + c[\dot{v}_{\textit{CAH}}+b\tanh{\frac{\dot{v}_{\textit{IDM}}-\dot{v}_{\textit{CAH}}}{b}}], & \: \textrm{otherwise}
  \end{array}
\right.
\end{equation}
where coolness factor $c$ is generally set to $0.99$.

There are two main transitions that IDM-Looming must handle:
\begin{itemize}
    \item From car-following behind the LA to negotiating with the MA
    \item From negotiating with the MA to car-following behind the MA
\end{itemize}

The model attempts to make these transitions relatively smooth by phasing in and out of a set of states:
\begin{itemize}
    \item If the TA is only following the LA with no other influences, use IDM-CAH with the LA.
    \item If a portion of the MA’s length is in the gap between the TA and LA, use both IDM-CAH with the LA and Looming-Mod with LA and MA, scaled based on the following:
\begin{equation}
\dot{v} = r\dot{v}_{\textit{IDM-CAH}} + (1-r)\dot{v}_{\textit{Looming-Mod}},
\end{equation}
        where $r$ = fraction of the MA's length not in the gap between the TA and LA.
    \item If the MA’s entire length is in the gap between the TA and LA, use both Looming-Mod with the LA and MA, and IDM-CAH with the MA, scaled based on the following:
\begin{equation}
\dot{v} = r\dot{v}_{\textit{Looming-Mod}} + (1-r)\dot{v}_{\textit{IDM-CAH}},
\end{equation}
        where $r$ = (the MA's lateral distance to the lane line) / (0.5 * lane width).
\end{itemize}

\subsubsection{MR-IDM}
The IDM-Looming model utilizes two different acceleration models to target both car-following and merging situations. To enable the use of a single, consistent acceleration model for both situations, the MR-IDM (Merge Reactive IDM) model was developed. This model uses IDM-CAH for the acceleration output, but simultaneously targets both the LA and any relevant MA, taking the maximum induced deceleration from the two actors. The only change to IDM-CAH is the input longitudinal distance ($s$): instead of the \emph{absolute} distance ($ds$), an \emph{effective} distance ($ds_e$) is calculated based on the trigonometric visual angle.

\begin{figure}
    \centering
\includegraphics[width=\linewidth]{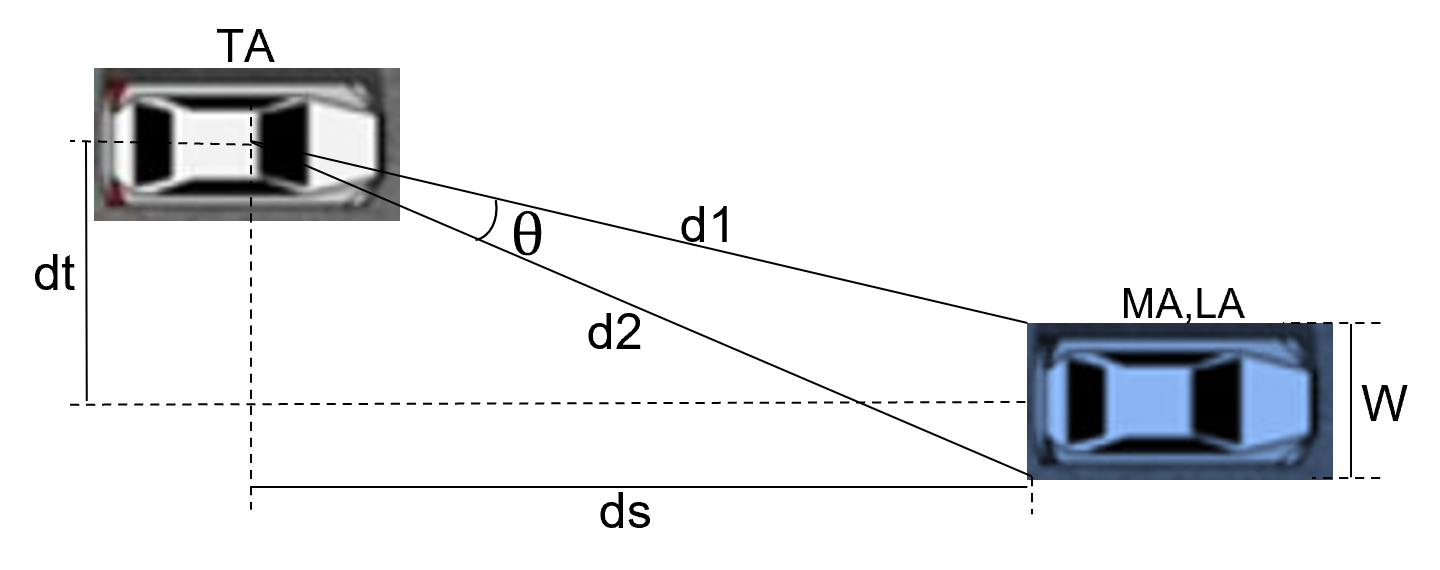}  
\caption{Visual angle $\theta$ from TA to a lead actor. The distances to the rear left and rear right of the leader are represented by $d_1,d_2$. $dt$ and $ds$ represent the relative longitudinal and lateral distances, respectively, from the TA driver's location to the rear of the leader.}
    \label{fig:visualangle}
\end{figure}

\begin{figure}
    \centering
\includegraphics[width=0.85\linewidth]{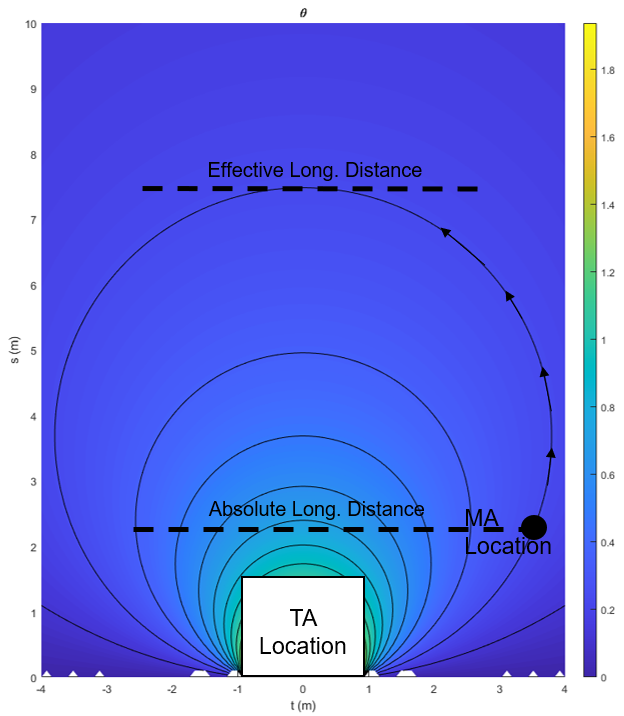}  
\caption{$\theta$ contour effect on effective distance.}
    \label{fig:dsavsdse}
\end{figure}

The looming-based models use an approximation of the visual angle $\theta$ shown in Fig. \ref{fig:visualangle}. A trigonometric interpretation of the visual angle can provide contours in the longitudinal/lateral relative distance space that relate to the TA’s varying reactivity to the MA, shown in Fig. \ref{fig:dsavsdse}. Any lateral relative position with a corresponding \emph{absolute} relative longitudinal distance has a certain visual angle to the rear of the actor ahead, and there exists an \emph{effective} longitudinal distance directly in front of the TA with the same visual angle. As the lateral distance approaches 0 meters (directly in front of the TA), the \emph{effective} distance approaches the \emph{absolute} distance. When using $ds_e$ instead of $ds$, the closer the MA moves laterally toward the TA, the stronger the deceleration reaction from the TA. Also, the TA’s deceleration can increase as the MA passes the TA, despite $ds$ growing larger, representing the MA's growing significance in the TA's field of view.

The effective distance is derived from $ds$, lateral distance ($dt$), and leading actor rear width ($W$) using the following equations:
\begin{equation}
d_1,d_2 = \sqrt{ds^2+(dt \pm W/2)^2},
\end{equation}
\begin{equation}
\theta = \cos^{-1}\frac{d_1^2+d_2^2-W^2}{2d_1d_2},
\label{eq:theta}
\end{equation}
\begin{equation}
ds_{e} = \frac{W(\frac{-(\cos{\theta}+1)}{\cos{\theta}-1})^{\frac{1}{2}}}{2}.
\label{eq:effdist1}
\end{equation}

Equations \eqref{eq:theta} and \eqref{eq:effdist1} can be simplified into Equation \eqref{eq:effdist2}:
\begin{equation}
ds_{e} = \frac{W}{2}\sqrt{\frac{(d_1+d_2)^2-W^2}{W^2-(d_1-d_2)^2}},
\label{eq:effdist2}
\end{equation}

This effective distance can be tuned such that the same lateral relative distance induces either a stronger or softer reaction from the TA using a new parameter: $\zeta$. This parameter scales the lateral relative distance used in the $d_1,d_2$ calculation. A $\zeta$ value below $1$ decreases the lateral distance, decreasing the effective distance and thus increasing the TA’s reactivity, while a value above $1$ has the opposite effect as 
\begin{equation}
dt_{new} = \zeta dt.
\label{eq:zetadt}
\end{equation}

\subsection{Parameter Optimization}
For a given traffic model, the identified TA from every valid real-world event was run through optimization. The goal was to minimize the fit error for each individual event by tuning the traffic model's parameters such that the simulation actor's speed time history matched that of the raw event. Theil's inequality coefficient \cite{theil1958economic} was used as the cost function. This coefficient is a measure of how well the simulation "forecast" represents the raw data. It is normalized between $0$ and $1$, providing a fair comparison between events of varying temporal length, as well as between traffic models, and it can capture both the time trend of the speed profile, as well as the value differences between simulation and raw data. The error is calculated using the following:
\begin{equation}
U = {\dfrac{\sqrt{\frac{1}{n}\sum_{i=1}^{n}(A_i-B_i)^2}}{\sqrt{\frac{1}{n}\sum_{i=1}^{n}A_i^2} + \sqrt{\frac{1}{n}\sum_{i=1}^{n}B_i^2}}} ,
\label{eq:theilscoefficient}
\end{equation}
where $A$ is the simulation speed profile, $B$ is the raw speed profile, and $n$ is the number of observations. This cost function was run through a nonlinear solver \emph{fminsearchbnd} using a MATLAB engine \cite{d8277fminsearchbnd} to provide reasonably quick optimization of bounded model parameters.
\begin{figure*}[h!]
    \centering
\includegraphics[width=0.95\linewidth]{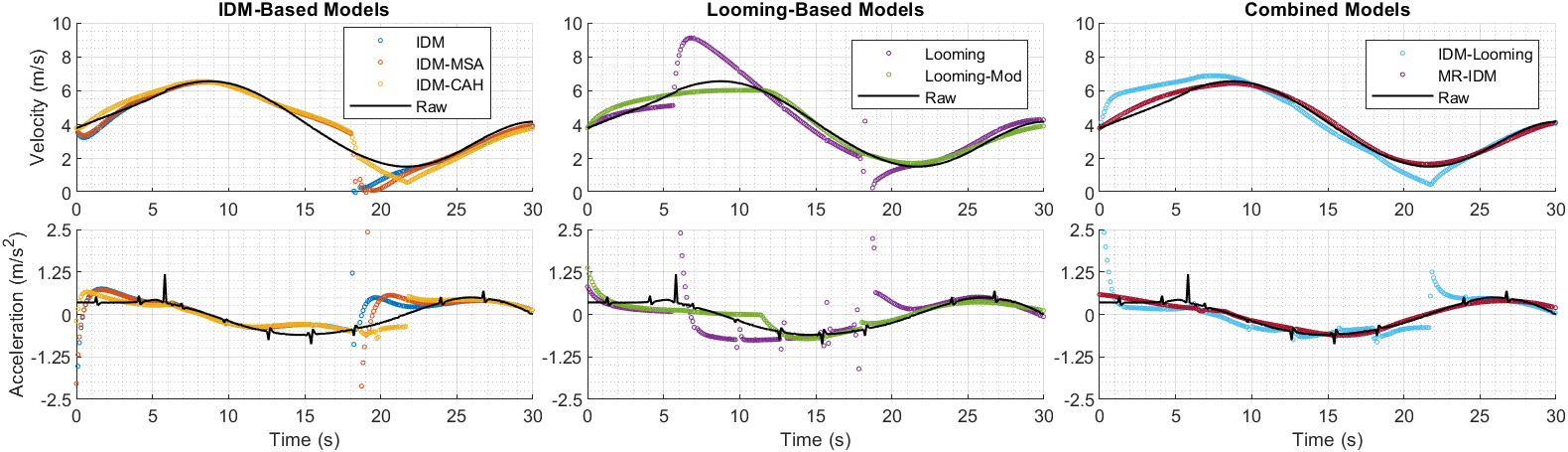}  
\caption{For this example event, the merge interaction occurs around $Time=6$-$18$ seconds. The larger drops in acceleration below $-2.5$ $\frac{m}{s^2}$ from IDM and IDM-MSA are not shown in the figure, while IDM-CAH was able to reduce the unreasonable deceleration. Note Looming-Mod's reduction of the looming model's acceleration jumps. Also, while IDM-Looming displayed discontinuities when transitioning in and out of the merge interaction, MR-IDM maintained smooth transitions and closely followed the raw speed profile.}
    \label{fig:errorresultsTraj}
\end{figure*}

 \section{Results and Analysis}\label{sec:behaveModel}
\begin{figure}
    \centering
\includegraphics[width=\linewidth]{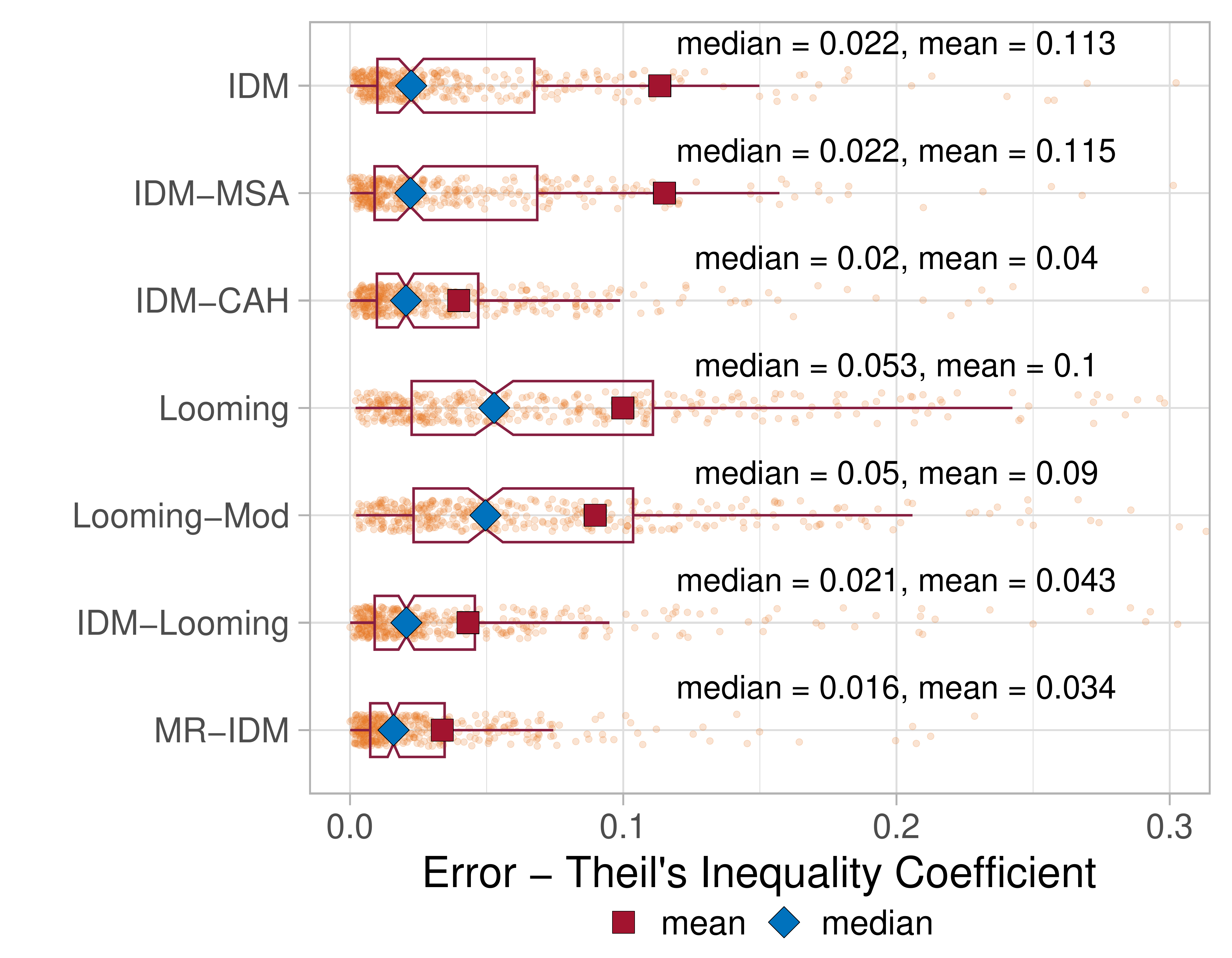}  
\caption{Theil's inequality coefficient results for model parameter optimization over $401$ events, focusing on the time range from the start of the merge interaction to $3$ seconds after the merge.}
    \label{fig:errorresultsbox}
\end{figure}

 The evaluation results of the Theil's inequality error of each proposed model on the real-world dataset are reported in Fig. \ref{fig:errorresultsbox}. The simulation result of the proposed models for one event from the real-world dataset is presented in Fig. \ref{fig:errorresultsTraj}. The base IDM was able to represent naturalistic behavior in simple car-following situations that were within the intended use case of the model, as well as low-density, large-gap merges; however, the model failed to replicate vehicle behavior during close merge negotiations. When the MA took over for the LA as the leader, this caused a discontinuity in the relative speed and gap size between vehicles, which in turn often caused unstable behavior. IDM-MSA improved over the base IDM by adding a lateral awareness component; however, it produced similar fits to the base IDM. The higher number of parameters reduced optimization performance, and $(\Delta{a_{\textit{MA}}} + p(\Delta{a_{\textit{TA}}}))$ was not suitably scaled to $\Delta$$a_{th}$, meaning a smooth leader transition from the LA to the MA was not guaranteed in the model logic. The IDM's instability with small longitudinal distances could also override the scaling factor to produce unreasonable decelerations. The IDM-CAH model reduced the IDM's unreasonable decelerations during leader changes, improving the IDM mean fit error by $65\%$ while reducing the standard deviation by $75\%$. IDM-CAH provides more traffic stability than  IDM and IDM-MSA, but still fails to acknowledge the MA until it becomes the TA's leader, causing late reactions in many instances.
  
The base looming model was able to specifically target the merge interaction by nature of its formulation but often failed to produce smooth transitions under the conditions outlined under Section \ref{sec:looming-mod}. This model also under-performed during strictly car-following interactions. The looming model's average error is lower than IDM and IDM-MSA due to more stability around the merge point, but poorer overall fits resulted in a worse median error. The Looming-Mod model reduced the base looming model’s acceleration jumps and improved upon the looming model's mean error by $10\%$. However, the other looming model weaknesses remained, including subpar car-following fits and the inability to slow down and create a larger gap for a faster MA.

IDM-Looming improved IDM's error mean by $62\%$ and standard deviation by $71\%$ and Looming-Mod's by $52\%$ and $39\%$ by combining IDM-CAH with Looming-Mod and restricting both models to their respective operational domains. The model generally produced smoother transitions; however, jumps in acceleration could still be seen at the point where car-following ends and merge interactions begin since, despite the transition logic, the two models remained uncoupled in their assumptions and parameters. The Looming-Mod model weaknesses also remained. Plus, this combination of models produced a large number of parameters to tune, reducing optimization performance and interpretability.

The fit result errors for the MR-IDM model surpassed those of the other models while removing acceleration jumps and decreasing the parameter count. MR-IDM improved upon IDM-CAH's mean and median error by $15\%$ and $22\%$, and IDM-Looming's by $22\%$ and $23\%$. Reliance on IDM-CAH for acceleration provided generally stable traffic. The looming-based model weaknesses were not seen due to the modified method for calculating the visual angle. Also, the reduced parameter set provided simpler optimization and more intuitive tuning of the model. Given an IDM car-following parameter set, a single parameter $\zeta$ could be used to tune the reactivity of the TA to an adjacent vehicle.

Since scalability in simulation was one of the goals of this work, the proposed models were also integrated into a high-fidelity simulation environment to simulate different highway traffic scenes to evaluate the performance of AV control and decision-making algorithms in a software-in-the-loop fashion. IPG-CarMaker was used to simulate the high-fidelity vehicle dynamics of a target car along with the road environment, while MATLAB-Simulink was used to simulate the traffic behavior models along with underlying vehicle controls. This setup was able to successfully simulate a highway merge scene involving 20 cars, each with their own traffic behavior model and underlying dynamics in real-time. This evaluation was performed on a laptop with a $64$-bit OS, Intel(R) Xeon(R) W-$2123$CPU @$3.6$ GHz  processor, and $16$ GB RAM. On closer inspection of the computation time of the traffic models themselves, it was found that the computation time was only about $0.5$ milliseconds but the CarMaker-Simulink interface proved to be a limitation to simulate up to $20$ cars. Given the model's low computation cost, it can also easily be exported to other simulation environments to simulate larger traffic streams for macroscopic evaluation of AV technology.

\begin{table}
\caption{Model feature comparison. "CF" indicates if there is a car-following component in the model. "$\theta$" indicates if the model uses a visual angle from the TA to MA. "Ramp" indicates if any ramp geometry is utilized. "LA" (Lateral Awareness) describes the ability of the model to acknowledge an adjacent, merging actor. "\# Params" tallies the number of model parameters.}
    \centering
\begin{tabular}{ |l||p{0.5cm}|p{0.5cm}|p{0.8cm}|p{0.5cm}|p{1.1cm}| }
 \hline
 Model & CF & $\theta$ &Ramp & LA & \# Params \\
 \hline
 IDM        & Yes & No  & No  & No  & 5 \\
 IDM-MSA    & Yes & No  & Yes & Yes & 7 \\
 IDM-CAH    & Yes & No  & No  & No  & 5 \\
 Looming    & No  & Yes & No  & Yes & 8 \\
 Looming-Mod& No  & Yes & No  & Yes & 5 \\
 IDM-Looming& Yes & Yes & No  & Yes & 10\\
 MR-IDM     & Yes & Yes & No  & Yes & 6 \\
 \hline
\end{tabular}
\end{table}
\section{Conclusion and Future Work}\label{sec:conclusion}
In this paper, we presented a few novel car-following models that are able to also model the reactive behavior of a main-lane traffic car to a merging car on the freeway. We compared the ability of these proposed models to represent real-world data by performing an evaluation on a unique real-world merging dataset that was collected as part of this work. The newly proposed MR-IDM model showed higher correlation with the real-world traffic data and also better accuracy and stability than the other models in our study. This model was also found to be fairly interpretable, tunable, and easily scalable for larger traffic simulations.

Future evaluation of the MR-IDM on other existing datasets will help ensure model generalizability and reduce any bias towards our training dataset. IDM-MSA is the only proposed model in this work to incorporate the proximity of the merge ramp end into TA's decision-making; thus, more focus can be placed on the effect of ramp geometry on traffic behavior during merge negotiations, as well as its integration into the more stable MR-IDM. Another area of future research is the effect of the looming angle's rate of change. As noted by \cite{wan2014modeling}, $\dot{\theta}_{\textit{MA}}$ has been cited as a dominant stimulus for traffic car-following behavior. More work can be done in incorporating this stimulus into the MR-IDM model, which currently relies on $\theta_{\textit{MA}}$. Also, this model’s scope can be extended beyond the merge interaction into other lane-change scenarios due to its general lateral reactivity.

\section{Acknowledgement}\label{sec:acknowledgement}
The authors would like to thank David Billet, Patrick Haley, Pawan Kallepalli, Shane McLaughlin, Joshua Radlbeck, Tyler Naes, Takayasu Kumano, Yosuke Sakamoto, and Paritosh Kelkar for their contribution to the overall project through its many phases, including planning, data reduction, software integration, and general discussions.

\bibliographystyle{IEEEtran.bst} 
\bibliography{reference/ref.bib}

\end{document}